# Automatic Intracranial Brain Segmentation from Computed Tomography Head Images

Bhavya Ajani, Bangalore, India.

*Abstract*— **Fast and automatic algorithm to segment Brain (intracranial region) from computed tomography (CT) head images using combination of HU thresholding, identification of intracranial voxels through ray intersection with cranium, special binary erosion and connected components per slice. Firstly, a thresholding is applied to create initial mask with voxels within desired HU range of soft tissues. Secondly, 'N' rays are projected outward from each mask voxel and fraction of rays intersecting with cranial voxels is taken as criteria to retain voxel as intracranial voxel. Thirdly, special binary erosion is applied to break connection between largest connected component representing brain and fragmented islands representing various cavities/sinuses. Lastly, only voxels belonging to largest connected component along both XY and YZ slices are retained as true intracranial brain voxels.**

## I. INTRODUCTION

Computed Tomography is preferred over other modalities like Magnetic Resonance for cranial imaging in acute stroke, intracranial hemorrhage and trauma. CT offers several advantages such as; low cost, wide availability, short imaging time and good resolution etc. However, CT scanning has limitations such as high dose of radiation, artifacts due to beam hardening etc.

In many of the post-processing applications like CT Neuro perfusion; brain is the primary area of interest and it is necessary to segment brain correctly from CT Head images to limit core processing to area of interest, improving performance. Unlike rich literature on brain segmentation from MRI data, research on segmenting brain from CT data is limited.

Fortunati V. et.al. [1] combines anatomical information based on atlas registration to segment various structures from CT Head and Neck data. Drawback of this approach is the computation time which could take up to an hour as per the author to fully segment a single CT Head & Neck data. In many of the clinical examination like Trauma or Stroke, time is critical and expectation is to complete examination within few minutes. Further, the results are sensitive to the selection of Atlas which may not represent anatomy for all patient age group.

Nevin Mohamed et. al. [2] proposed a modified fuzzy C-Mean based segmentation of Brain from CT data. However, this approach is sensitive to the presence of noise in the data and pathology. Further, it does not differentiate between soft tissues from intracranial and extracranial regions.

Hu Q. et. al. [3] proposed an improved fuzzy C-Mean based threshold and mask propagation approach to segment brain from CT data. The algorithm is not fully automatic as it requires a reference image.

Masatoshi Kondo et. al. [5] have proposed an automated brain tissue extraction algorithm which uses smallest distance of a voxel from skull as criteria to classify brain tissue which is comparable to 'ray intersection with cranium criterion' used in this approach at step 2. However, the stated approach has following limitations:

  a. For any voxel if all distances are not measurable the voxel is not considered as part of brain tissue. This could create long strips of false negatives along area where voxels falls in line of sight of a foramen/cavity.
  b. No special techniques are applied to remove areas of false positives representing cavities/sinuses.

The overall method proposed here is superior as it compensates for above two source of error and this is validated by the fact that the rate of false-negatives and false-positives are substantially lower as compare to the method proposed by Kondo et. al. Further, method proposed here is 30 times faster as it does away with multiple erosions and template match.

## II. METHODOLOGY

CT has two unique advantages which can be leveraged for brain segmentation. First, for any calibrated CT the HU range of cranial (bony skull) region and soft tissues has no overlap thus segmenting bony skull area from soft tissue becomes trivial using a simple threshold. This threshold may be applied heuristically or estimated using any standard binarization algorithm like Otsu or derived through clustering method or histogram based techniques. Second, anatomically brain is enclosed within bonny cranium and as bones can be identified easily, this anatomical enclosure can be utilized as criteria to identify voxels belonging to brain (Gray/White matter) i.e. intracranial.

The coordinate system (xyz) of a volumetric dataset is represented according to the standard radiological convention: x runs from subject's right to left, y from anterior to posterior, and z from superior to inferior. The intensity of a voxel (x, y, z) in original data is denoted as I(x, y, z) whereas corresponding label in binary brain mask is denoted as M(x,y,z).

For a mask voxel Mask(x,y,z) a XY slice is defined as slice comprising of all voxels with same 'z' co-ordinate. YZ slice is defined as slice comprising of all voxels with the same 'x' co-ordinate. XZ slice is defined as slice comprising of all voxels with the same 'y' co-ordinate.

The method proposed here can be outline as shown in flow diagram, Fig. 1.

### A. Create Initial Brain Mask

Using lower HU threshold for air (lowThres) and upper HU threshold for bones (highThres), threshold original CT data to get an initial binary mask M(x,y,z). Here, lower HU threshold is taken as -40 HU and upper HU threshold is taken as 160 HU heuristically.

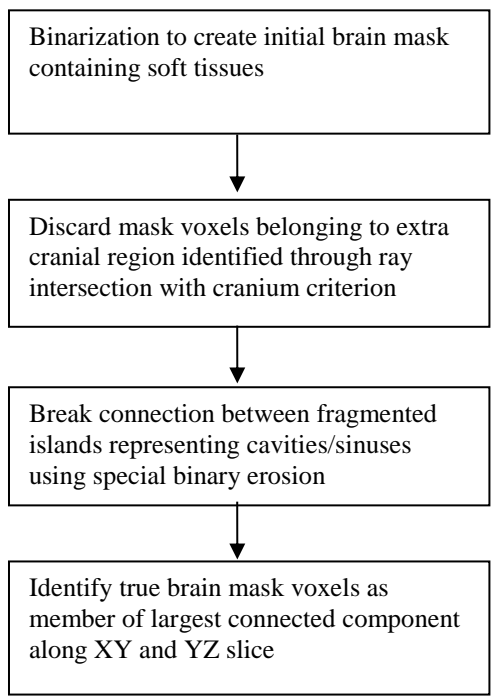

Fig.1. Flow diagram for proposed method on segmenting brain from CT images.

$$Mask(x,y,z) = \begin{cases} 1 & lowThres <= I(x,y,z) <= highThres \\ 0 & otherwise \end{cases}$$

The lower and upper HU threshold may be taken heuristically or can be determine from the original data itself using any standard techniques like clustering or histogram.
At end of this step we get a mask consisting of soft tissue voxels belonging to both intra and extra cranial regions.

### B. Discard Mask voxels belonging to extra cranial region

Anatomically brain is surrounded by protective bony cranium which has large HU. Hence, theoretically if we project multiple rays from a voxel belonging to brain region majority of them should hit bony cranium. This fact is utilized in this step to select only those mask voxels, for which multiple rays when traced out intersect bony region identified as voxel with HU above some lower threshold. Here HU of 300 is taken as lower threshold to identify cranial voxels. This lower threshold for bone again can be set heuristically or can be determined from the original data itself using technique such as clustering or histogram.

Here, total of 8 rays were projected away from the mask voxel on axial plane at an angle of 0°, 45°, 90°, 135°, 180°, 225°, 270°, 315°. If 7 or more rays hit bony cranium the voxel is consider as intra cranial and thus retained otherwise it is discarded as extra cranial voxel. All 8 rays are not considered because for some intracranial voxels one of the rays may not intersect cranium but pass through various foramen or canals. Fig. 2 shows rays projected outward from one of the mask voxel hitting cranium.

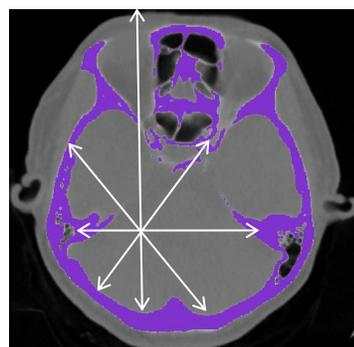

Fig. 2. Projected rays intersecting with cranium (magenta) for one of the candidate mask voxel.

At the end of this step the brain mask have single largest connected component representing true brain from intracranial region and few fragmented islands representing cavities/sinuses from extracranial region as these are enclosed within bony structure. Also mask at this stage contain sparse noisy false positive voxels. Fig. 3 shows the mask at the end of this step.

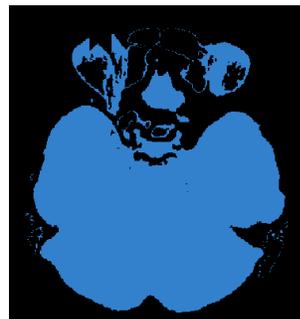

Fig. 3. Segmented brain mask after discarding extra cranial voxels.

## C. Break connection between fragmented islands and smoothen brain mask using special binary erosion

To remove any noisy false positive voxels from mask and to reduce connection between largest connected component representing the desired brain mask and fragmented islands coming from extracranial cavities/sinuses a special binary erosion is performed on the mask coming from previous step. Special erosion flip any true voxel mask if it does not have contiguous true voxels along either horizontal 'x' or vertical 'y' direction for minimum distance of 10 mm. This operation removes all noisy false positives from the mask as these components are never larger than 10 mm in extreme direction and also opens up any narrow channel of connection between the single largest connected component representing true brain mask and fragmented islands coming from cavities/sinuses.

This special erosion is applied twice. Specialized binary erosion offers three distinct advantages as compare to binary opening operation to remove narrow connection. (a) The results are not dependent upon the selection of structural element. (b) The brain mask edge along surface is preserved. (c) Operation is much faster compare to binary opening as this can be optimized as run length encoding problem and fact that binary opening require two morphological operations, erosion followed by dilation.

At the end of this step we have a mask with single largest connected component representing desired brain voxels along with few fragments coming from cavities/sinuses; some of which are isolated completely from the main brain mask in 3D whereas some are still not isolated completely in 3D but are isolated when viewed on 2D XY, YZ or XZ slices independently. In next step we shall utilized this distinctiveness to extract true brain mask. Fig. 4 shows mask at the end of this step. Compare the same with Fig. 3.

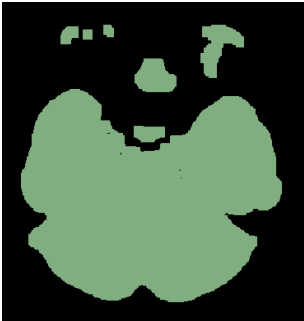

Fig. 4. Segmented brain mask after applying special binary erosion.

## D. Identify true brain mask voxels as member of largest connected component along XY and YZ slice

In the final step, we only retain mask voxels if it belongs to largest connected component along both XY and YZ slices. For 3D mask data we identify 2D largest connected component along all XY slices for various values of 'z' represented as $LCC_{xy}(z)$. Similarly, we identify 2D largest connected component along all YZ slices for various values of 'x' represented as $LCC_{yz}(x)$. We retain mask voxel $M(x,y,z)$ only if it belongs to both $LCC_{xy}(z)$ & $LCC_{yz}(x)$.

$$Mask(x,y,z) = \begin{cases} 1 & LCC_{xy}(z) \ \& \ LCC_{yz}(x) \\ 0 & \text{otherwise} \end{cases}$$

Finding largest connected component in 3D necessary may not give desired brain mask as few of the fragmented islands representing cavities/sinuses from extra cranial regions are still connected to brain mask along one direction due to large slice thickness in CT data. Hence, these fragmented islands can be removed by checking if voxels belongs to largest connected component per slice basis for any 2 principal planes, here XY & YZ. Fig. 5 shows final brain mask at the end of this step.

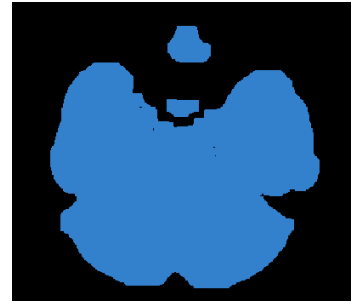

Fig. 5. Segmented brain mask at the end of algorithm.

## III. RESULTS

The algorithm was implemented in C++ using ITK on Intel Core i7-4790 CPU @ 3.60 GHz with 16 Gb RAM. Results were quantitatively validated against 1 non-enhanced CT data set (512x512x45) voxels with spacing of 0.488 mm/0.488 mm/2.5 mm.

The brain mask was extracted in 1.1 seconds. The binary ground truth mask was created by manually drawing brain contours along axial slices in Slicer tool with 'true' voxel representing intra-cranial brain tissue. False negative rate (FN) and False positive rate (FP) were used to quantify the accuracy of segmentation. False negative rate (FN) is the percentage of voxels marked as false on segmented brain mask but true on ground truth against total true voxels on ground truth. False positive rate (FP) is the percentage of voxels marked as true on segmented brain mask but false on ground truth against total true voxels on ground truth.

$$False\ Negative = 100 \times (!BM \cap GT)/GT \quad (1)$$

$$False\ Positve\ = 100 \times (BM \cap !GT)/GT \quad (2)$$

Where, BM is the binary brain mask extracted by algorithm and GT is the binary ground truth.

For the normal CT data the computed FN and FP were 2.31% and 0.9% respectively which indicate accurate segmentation. Fig. 6 & 7 shows segmented brain mask overlay over original CT data. Further, the algorithm was validated with same CT data simulated with 15° rotation along [1,-1,1] axis vector. Fig. 8 shows segmented brain mask overlay over simulated CT data.

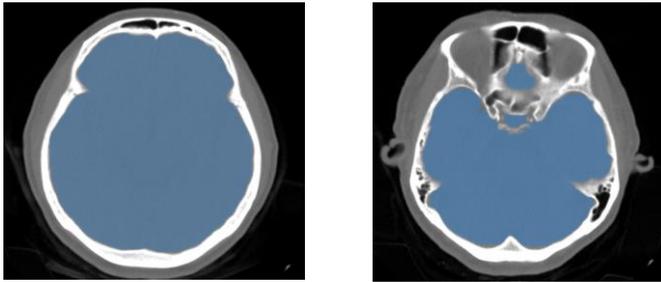

Fig. 6. Brain segmentation mask shown in light blue overlay over original CT data along Axial plane.

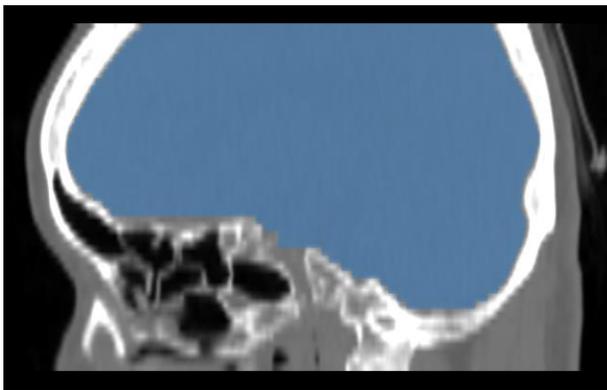

Fig. 7. Brain segmentation mask shown in light blue overlay over original CT data along Sagittal plane.

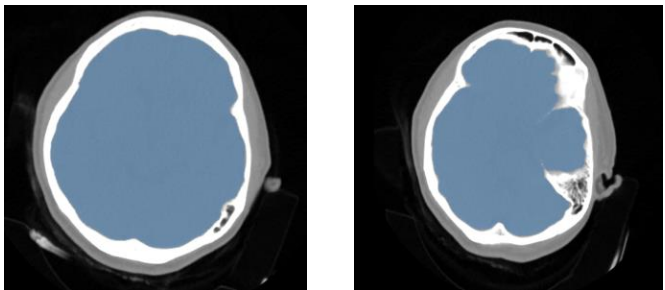

Fig. 8. Brain segmentation mask shown in light blue overlay over simulated CT data with rotation along Axial plane.

## IV. DISCUSSION

As results indicate, algorithm is comparatively fast as compare to other extraction algorithm using template atlas or Fuzz-C Mean clustering. Further, algorithm is robust enough to work even with rotation of the head. Small value of False positive rate and relative large value of False negative rate indicate under segmentation specially along the inferior convexity of temporal lobe. This can be improved by considering 2D largest connected component along XZ slice along with proposed XY & YZ slice. However, these shall reduce false negative rate at the expense of increasing false positive rate and shall increase computation time.

Various thresholds for soft tissue and bone classification were set heuristically due to well defined HU range for various anatomical tissues and fact that HU range of soft tissue and bones has large separation. However, scope exists to improve threshold estimation further using any techniques like clustering from data itself.

Results could be further improved by increasing the number of rays projected outward from a voxel to determine intracranial voxel. Various strategies could be implemented to select candidate for intracranial voxel based on number of rays intersecting cranium or distance to cranium.

## V. CONCLUSION

Quick and robust algorithm to extract brain mask from CT data is proposed to improve performance of clinical applications focusing on brain as area of interest and to improve diagnosis.

An algorithm utilizing the information that brain is enclosed within cranium and thus rays traced outward from brain voxel shall intersect cranium is proposed. To improve results, special binary erosion is applied followed by slice by slice membership to largest connected component along at least two principal plane is added. As algorithm consists of simple operations it is fast. Further, as key features in extracting brain mask are topological rather than statistical, the algorithm is robust to noise, presence of pathology and patient age.